\documentclass[12pt]{article} 
\usepackage[sectionbib]{natbib}
\usepackage{array,epsfig,fancyheadings,rotating}
\usepackage[]{hyperref}  
\usepackage{sectsty, secdot}
\sectionfont{\fontsize{12}{14pt plus.8pt minus .6pt}\selectfont}
\renewcommand{\theequation}{\thesection\arabic{equation}}
\subsectionfont{\fontsize{12}{14pt plus.8pt minus .6pt}\selectfont}

\usepackage{floatrow}
\usepackage{float}
\usepackage{amsmath}
\usepackage{amssymb}
\usepackage{amsfonts}
\usepackage{amsthm}
\usepackage{natbib}
\usepackage[]{algorithm}
\usepackage{algpseudocode}
\usepackage{array}
\usepackage{threeparttable}
\usepackage{multicol,multirow}
\usepackage{tabularx}
\usepackage{graphicx}
\usepackage{makecell}
\usepackage{caption}

\theoremstyle{definition}

\begin{document}
\pagenumbering{arabic}


\renewcommand{\baselinestretch}{2}

\markright{ \hbox{\footnotesize\rm 
}\hfill\\[-13pt]
\hbox{\footnotesize\rm
}\hfill }

\markboth{\hfill{\footnotesize\rm Scott H. Koeneman and Joseph E. Cavanaugh} \hfill}
{\hfill {\footnotesize\rm Bootstrap Goodness-of-Fit} \hfill}

\renewcommand{\thefootnote}{}
$\ $\par


\fontsize{12}{14pt plus.8pt minus .6pt}\selectfont \vspace{0.8pc}
\centerline{\large\bf A New Bootstrap Goodness-of-Fit Test for}
\vspace{2pt} 
\centerline{\large\bf Normal Linear Regression Models}
\vspace{.4cm} 
\centerline{Scott H. Koeneman and Joseph E. Cavanaugh} 
\vspace{.4cm} 
\centerline{\it Thomas Jefferson University and the University of Iowa}
 \vspace{.55cm} \fontsize{9}{11.5pt plus.8pt minus.6pt}\selectfont


\begin{quotation}
\noindent {\it Abstract:}
In this work, the distributional properties of the goodness-of-fit term in likelihood-based information criteria are explored. These properties are then leveraged
to construct a novel goodness-of-fit test for normal linear regression models that relies on a non-parametric bootstrap. Several simulation studies are performed to
investigate the properties and efficacy of the developed procedure, with these studies demonstrating that the bootstrap test offers distinct advantages as compared
to other methods of assessing the goodness-of-fit of a normal linear regression model.

\vspace{9pt}
\par
\end{quotation}\par

\def\thefigure{\arabic{figure}}
\def\thetable{\arabic{table}}

\renewcommand{\theequation}{\thesection.\arabic{equation}}

\fontsize{12}{14pt plus.8pt minus .6pt}\selectfont

\section{Introduction and Background}

    	Broadly, the term \textit{goodness-of-fit} as it pertains to statistical modeling refers to the degree to which a certain model
		and its associated assumptions align with the observed data. Assessments of this property may take the form of visual diagnostics, quantitative diagnostics, or formal hypothesis tests to help
		determine whether the assumptions of a given model are met.

		The notion of goodness-of-fit is relevant in the model selection realm of information criteria. Using $\ell(\hat{\theta}|y)$ to denote the log-likelihood of the fitted model and $p$ to denote the
		number of estimated parameters, the \textit{Akaike information criterion} (AIC) for a fitted model can be expressed as
		\begin{equation*}
			AIC = -2 \ell(\hat{\theta}|y) + 2 p.
		\end{equation*}
		The $-2 \ell(\hat{\theta}|y)$ term based on the empirical log-likelihood is known as the goodness-of-fit term as it represents the degree to which the fitted model
		conforms to the observed data $y$ (\cite{Cavanaugh}). This term is also present in other likelihood-based information such as the \textit{Bayesian information criteria} (\cite{Schwarz}). The goodness-of-fit
		term will only grow smaller, indicating better conformity to the data, as complexity is added to the model, and thus while it is a reasonable measure of goodness-of-fit, other considerations are needed to avoid overfitting.

		Linear regression analysis makes a number of assumptions about the data at hand. Perhaps the most basic assumption is that each outcome is independent given the set of predictors at hand.
		Additionally, it is assumed that the outcomes, conditioned on the covariates, have a mean that is a linear function of the covariates, with this property being referred to as \textit{linearity},
		with errors that are normally distributed with a mean of zero. These errors are assumed to be independent and identically distributed with a constant variance across all observations, with this
		constant variance property being referred to as \textit{homoskedasticity} (\cite{Kutner}). While this work will focus on assessments of the goodness-of-fit of traditional normal linear regression models, it is worth noting
		that assessments of goodness-of-fit exist for many other modeling paradigms as well, and many of the same principles apply.
		
		One method of visually assessing the assumptions of linearity and homoskedasticity when using linear
		regression is to plot the residuals against either their corresponding predictor values or the corresponding observed values of the outcome, with these figures being called \textit{residual plots} (\cite{Miles}). Across different values
		of the predictors and observed values, the residuals should not exhibit any specific pattern other than a mean of zero and constant variance if the assumptions of the linear regression model are met.
		Thus, deviations from this pattern in the form of a residual plot that exhibits curvature, or a change in spread, can indicate violations
		to linearity and homoskedasticity, respectively. 

		However, this method of visual inspection carries with it certain limitations. When there are a large number of predictors present in a model, it may be impractical to visually inspect every
		possible residual plot with any degree of scrutiny. Additionally, there may be patterns in deviations that are undetectable upon visual inspection, such as a curved pattern that is less noticeable
		due to the scale of the predictors. If \textit{heteroskedasticity}, meaning the absence of homoskedasticity, is induced by covariates that have not been observed, a visual inspection will likely
		not reveal this violation of assumptions.

		Hypothesis tests for the goodness-of-fit of a linear model offer an alternative to visual methods. These tests posit a null hypothesis that the model adequately accommodates the data and that
		the associated assumptions are satisfied, with an alternative that the assumptions are violated in some fundamental way. One such test is the \textit{Breusch-Pagan test}, which posits a null hypothesis
		that a linear regression model does not violate homoskedasticity (\cite{Breusch}). This test involves performing an auxiliary linear regression on a transformation of the squared residuals from
		the candidate linear regression model against the covariates of interest. Essentially, the more of the variation in the squared residuals that is explained by the auxiliary regression, the more one suspects violations to homoskedasticity.

		However, one limitation of the Breusch-Pagan test is that it is only designed to detect a relationship between the covariates and the squared residuals that is linear, and thus it will not
		produce efficacious results if the heteroskedasticity present is not linear (\cite{Waldman}). An alternative to the Breusch-Pagan test in this matter is the \textit{White test} for homoskedasticity,
		which shares the same null hypothesis of homoskedasticity of a linear model as the Breusch-Pagan test (\cite{White1980}). The White test is similar to the Breusch-Pagan test in that it involves
		an auxiliary regression with the squared residuals as the outcome, but here, the regressors are all of the covariates in the original fitted model in addition to their squares and cross
		products. This produces a test statistic that is sensitive to deviations to the null hypothesis in the form of
		heteroskedasticity related to the squares and cross products of regressors. Additionally, the test may indicate misspecification of the model if the cross products of certain
		regressors should be included in the model, but are not.

		The various ways in which the White test can detect misspecification bring to attention a weakness of the hypothesis test methods as opposed to methods of visual inspection.
		When a hypothesis is rejected, we can be reasonably confident that an assumption is violated; however, by simply performing the test, we do not glean much information as to how exactly the
		model may be misspecified. In the case of rejection of the White test, one may not know if we reject because cross-products should be included as covariates, or if perhaps there is heteroskedasticity
		present related to one of the covariates already present. This is in contrast to the method of visually observing residual plots wherein specific issues may be easier to identify.
		If a plot of the residuals versus a covariate exhibits a curved pattern, one may posit a new model that accounts for this curved relationship (\cite{Miles}).
		However, hypothesis tests offer little direction other than that the current model exhibits lack-of-fit in some manner, and further investigation is necessary to determine next steps.

		In addition, both the White and Breusch-Pagan tests cannot detect heteroskedasticity induced by unobserved covariates, as their auxiliary regressions only use what has been observed. Such heteroskedasticity
		may not affect the bias and consistency of effect estimates or the validity of inference, but may lead to a loss of efficiency of estimates. 

		When employing likelihood theory to obtain parameter estimates for a model, one can often leverage the properties of maximum likelihood estimators (MLEs) to produce reasonable estimates of the variance of the
		estimators, and thus perform inference (\cite{Millar}). However, these properties are only guaranteed to hold when the model is properly specified, and improper or inefficient estimates may result when assumptions
		do not hold. This has given rise to \textit{robust variance estimators} that rely on fewer assumptions than classical likelihood theory, yet can still quantify the variability of a statistic
		at hand.

		Huber first provided justifications for consistency and asymptotic normality of maximum likelihood estimators under conditions weaker than had been previously shown (\cite{Huber}). White
		expanded upon this notion by deriving covariance matrix estimates for maximum likelihood estimators that are robust to a variety of different types of misspecification and only assume
		conditional independence and certain other regularity conditions (\cite{White1980}). Defining $\theta_*$ as the \textit{pseudo-true} parameter and $\hat{\theta}_n$ as the MLE for a sample of size
		$n$, White showed the asymptotic relation
		\begin{equation*}
			\sqrt{n} (\hat{\theta}_n - \theta_*) \xrightarrow[]{d} N(0, C(\theta_* ) ) .
		\end{equation*}
		The large sample covariance matrix $C(\theta)$ can be defined using the matrices 
		\begin{equation*}
			A(\theta) = E \left[ \frac{\partial^2 f(y_t,\theta)}{\partial \theta_i \partial \theta_j} \right] 
		\end{equation*}
		and
		\begin{equation*}
			B(\theta) = E \left[ \frac{\partial f(y_t,\theta)}{\partial \theta_i} \frac{\partial f(y_t,\theta)}{\partial \theta_j} \right] 
		\end{equation*}
		where $i = 1,...,p$ and $j = 1,...,p$ respectively. These matrices then combine to define $C(\theta)$ as
		\begin{equation*}
			C(\theta) = A(\theta)^{-1} B(\theta) A(\theta)^{-1} ,
		\end{equation*}
		and thus evaluating this quantity at $\theta_*$, one arrives at the robust asymptotic variance estimate. The resulting estimator, and those that are similar in form, are often called \textit{sandwich}
		variance estimators due to one quantity being sandwiched in between two identical others to form the statistic.

		As in general one will not know the pseudo-true parameter $\theta_*$, White indicates that the matrices
		\begin{equation*}
			A_n(\theta) = \frac{1}{n} \sum_{t=1}^{n} \frac{\partial^2 f(y_t,\theta)}{\partial \theta_i \partial \theta_j}
		\end{equation*}
		and
		\begin{equation*}
			B_n(\theta) = \frac{1}{n} \sum_{t=1}^{n} \frac{\partial f(y_t,\theta)}{\partial \theta_i} \frac{\partial f(y_t,\theta)}{\partial \theta_j} 
		\end{equation*}
		can be calculated from the data and evaluated at the MLE to form
		\begin{equation*}
			C_n(\hat{\theta}) = A_n(\hat{\theta})^{-1} B_n(\hat{\theta}) A_n(\hat{\theta})^{-1} .
		\end{equation*}
		It can then be shown that 
		\begin{equation*}
			C_n(\hat{\theta}_n) \xrightarrow[]{a.s.} C(\theta_* ) .
		\end{equation*}
		Thus, we have a variance estimator for the MLE that is both robust to much misspecification and can be calculated using the data, but yet also will be approximately equivalent to the standard
		likelihood theory estimator if the model is correctly specified. This sandwich estimator, and others like it, can be used to perform inference related to the parameters if one calls into question
		the strong assumptions involved in using the traditional maximum likelihood estimator. However, if these assumptions cannot be met and the model appears to be misspecified, one may ponder
		the merit of performing inference on the pseudo-true parameters in the first place (\cite{Huber}). Therefore, robust variance estimators serve as a hedge against slight deviances from a model
		being correctly specified, not as a tool that remediates poor model selection.

\section{Derivations and Test Formulation}

		We will first explore the variance of the log-likelihood goodness-of-fit term present in likelihood-based information criteria. We will assume that we are in a
		scenario where a normal linear regression model is being fit to the data of interest, and that this model is not misspecified. Thus, this model is of the proper
		parametric family and contains the requisite mean structure, although the mean structure may contain extraneous variables in the
		case of an overspecified model.

		Assuming a linear model has been fit using maximum likelihood with fitted parameters $\hat{\theta}$, the goodness-of-fit term can be decomposed as
		\begin{equation}
			-2 \ell (\hat{\theta}  ) = n \log(2 \pi) + n + n \log(\hat{\sigma}^2 ) ,
		\end{equation}
		where $\hat{\sigma}^2$ denotes the maximum likelihood estimate for the error variance $\sigma^2$. Note that the only term here that is random is
		the $n \log(\hat{\sigma}^2)$ term. Thus, if we can quantify the variability of this term, we can quantify the variability of the entire goodness-of-fit
		term.

		To achieve this end, we first consider $\hat{\sigma}^2$. As the linear model is assumed to not be misspecified, and $\hat{\sigma}^2$ is a maximum likelihood
		estimator, this estimator will have an asymptotic variance related to the inverse of the Fisher information (\cite{Fisher}). In the case of a single
		observation, the Fisher information as it relates to the parameter vector $\theta' = [\beta', \sigma^2]$, where $\beta$ represents the regression coefficients
		present in the model, can be shown to be
		\begin{equation*}
			- E \left[ \frac{\partial^2 \ell (\hat{\theta}  )}{\partial \theta^2} \right] = \mathcal{I}_{n}(\theta) =
			\begin{bmatrix}
				\frac{X' X}{\sigma^2} & 0 \\
				0 & \frac{n}{2 \sigma^4} \\
			\end{bmatrix}
			,
		\end{equation*}
		where $X$ is the design matrix of the regression, and $0$ is a vector of zeroes. Thus, if we assume we have a single observation
		such that $n=1$,we may take the inverse of the Fisher information matrix and isolate the element related to the error variance $\sigma^2$. We see that
		this element will be
		\begin{equation*}
			\mathcal{I}_{1}^{-1}(\sigma ^2) = 2 \sigma ^4 .
		\end{equation*}
		
		Using the above relation, and applying the property of asymptotic normality of the MLE $\hat{\sigma}^2$ in this case of a properly specified normal linear
		regression model, we see that the asymptotic distribution
		\begin{equation*}
			\sqrt{n} (\hat{\sigma}^2 - \sigma^2) \xrightarrow[]{d} N(0, 2 \sigma ^4 )
		\end{equation*}
		should hold.

		To find the variance of $-2 \ell (\hat{\theta}  )$, we must find the variance of $n \log(\hat{\sigma}^2)$.
		Additionally, the above asymptotic distribution involves the true $\sigma^2$ to which we will not have access in most modeling scenarios.

		We will solve both of these issues by employing the delta method (\cite{Rao}). We propose a transformation of the form
		\begin{equation*}
			g(x) = \log(x) .
		\end{equation*}
		Thus, applying the delta method to our above asymptotic distribution with $g(x)$ as the function of interest, we see that
		\begin{equation*}
			\sqrt{n} ( \log (\hat{\sigma}^2) - \log(\sigma^2)) \xrightarrow[]{d} N(0, 2) .
		\end{equation*}

		Armed with the above asymptotic relationship and assuming that this asymptotic property approximately holds in a setting with a finite $n$, we have that
		\begin{equation*}
			n\log(\hat{\sigma}^2) \; \dot\sim \; N \left( n\log(\sigma^2), 2n \right) .
		\end{equation*}
		Thus, assuming that the model is appropriately specified, the variance of $n\log(\hat{\sigma}^2)$ will be approximately $2n$. Applying this variance back to the goodness-of-fit term,
		we see that the approximation
		\begin{equation*}
			Var \left[ -2 \ell (\hat{\theta}  ) \right] \approx 2n
		\end{equation*}
		is justified. Furthermore, $2n$ could also serve as an approximation to the variance of AIC or BIC for this correctly specified linear regression model. This approximation
		is extremely easy to compute, and has the same form no matter the complexity of the design matrix $X$ or value of the true parameters $\theta$, making it useful as a
		general tool.

		It should be noted that this approximation does rely on asymptotic properties. In the Appendix, an exact variance for $-2 \ell (\hat{\theta})$ is found which does
		not rely on asymptotic properties. However, this variance is more complicated to compute than the simple approximation $2n$, and was not found to provide any meaningful
		benefit over $2n$ when used in procedures developed later in this work.

		With the previous derivation in hand, we now develop an estimator for the variance of $n\log(\hat{\sigma}^2)$, and thus $-2 \ell (\hat{\theta})$,
		that need not assume a given fitted normal linear regression model is correctly specified.

		Assume that we once again fit a linear regression model with parameters $\theta' = [\beta', \sigma^2]$ to our data, and suppose we do not know whether this model is correctly specified.
		We wish to construct an estimator for $Var \left[ -2 \ell (\hat{\theta}  ) \right]$. This estimator will be constructed using the White robust sandwich variance estimator, employing
		much of the notation related to this development that was introduced in the first section of this work (\cite{White1980}).

		We let $I_{n} (\theta)$ denote the observed information matrix with regards to our specified linear regression model. With $X$ denoting the $n$ by $r$ design matrix, we see that
		the quantity $A_n (\theta)$ used in the White estimator is found to be 
		\begin{equation*}
			A_n(\theta) = \frac{1}{n}
			\begin{bmatrix}
				\frac{X'X}{\sigma^2} & \left[ \frac{-(y-X\beta)'X}{\sigma^4} \right]' \\
				\left[ \frac{-(y-X\beta)'X}{\sigma^4} \right] &  \frac{n}{2 \sigma^4} - \frac{(y-X\beta)'(y-X\beta)}{\sigma^6}
				\end{bmatrix}
				= -\frac{1}{n} I_n(\theta) .
		\end{equation*}
		Now let $x_i$ be the $i$th row of the design matrix $X$, and $y_i$ be the $i$th observation of the observation vector $y$. With these constructs at hand, we can define the score components
		for individual observations in our sample as
		\begin{equation*}
			U_i(\theta) = 
			\begin{bmatrix}
				\frac{(y_i-x_i \beta)x_i'}{\sigma^2} \\
				\frac{-1}{2 \sigma^2} + \frac{(y_i - x_i \beta)^2}{2 \sigma^4}
			\end{bmatrix}
			.
		\end{equation*}
		With these score components defined, we can then use the preceding to represent the matrix $B_n (\theta)$ used in the White estimator as
		\begin{equation*}
			B_n(\theta) = \frac{1}{n} \sum_{i=1}^{n} U_i(\theta) U_i(\theta)' .
		\end{equation*}

		Thus, with $B_n(\theta)$ and $A_n(\theta)$ defined, these matrices can be evaluated at the maximum likelihood estimator $\hat{\theta}$ and be used to define
		\begin{equation*}
			\begin{split}
				C_n(\hat{\theta}) & = A^{-1}_n(\hat{\theta}) B_n(\hat{\theta}) A^{-1}_n(\hat{\theta}) \\
				& = n I_n^{-1}(\hat{\theta}) [\sum_{i=1}^{n} U_i(\hat{\theta}) U_i(\hat{\theta})'] I_n^{-1}(\hat{\theta}) ,
			\end{split}
		\end{equation*}
		where $C_n(\hat{\theta})$ will be a $r+1$ by $r+1$ matrix that can be used as an estimator of the asymptotic variance-covariance matrix of $\hat{\theta}$. This estimator is robust to misspecification
		of the model. Consider the bottom rightmost element of this matrix, that being the $(r+1)^{th}$ element of the $(r+1)^{th}$ column of the matrix. This element will correspond to the large-sample
		robust variance of $\hat{\sigma}^2$. Let $s(\theta)$ refer to this corresponding element in the case of the theoretical matrix $C(\theta)$, and $s_n(\hat{\theta})$ refer to this
		corresponding element of the estimator $C_n(\hat{\theta})$. By White's results, it can then be seen that
		\begin{equation}
			\sqrt{n} (\hat{\sigma}^2 - \sigma_*^2) \xrightarrow[]{d} N(0, s(\theta_*)) ,
		\end{equation}
		where $\sigma_*^2$ denotes the pseudo-true parameter related to $\sigma^2$ in the case of potential misspecification.

		We will again employ the delta method to transform the asymptotic distribution to a form that is more suitable. We once more define a transformation of
		\begin{equation*}
			g(x) = \log(x) .
		\end{equation*}
		Applying this transformation to the asymptotic distribution presented in (2.2), we arrive at the relation
		\begin{equation*}
			\sqrt{n} ( \log (\hat{\sigma}^2) - \log(\sigma_*^2)) \xrightarrow[]{d} N \left( 0, \frac{1}{\sigma_*^4} s(\theta_*) \right) .
		\end{equation*}
		Using this asymptotic distribution, one can arrive at an approximate distribution for $n\log(\hat{\sigma}^2)$ as
		\begin{equation*}
			n\log(\hat{\sigma}^2) \; \dot\sim \; N \left( n\log(\sigma_* ^2), \frac{n}{\sigma_*^4} s(\theta_*) \right) ,
		\end{equation*}
		which could be suitable for use in finite sample sizes that are sufficiently large. However, $\sigma_*^2$ and $s(\theta_*)$ are likely to be unknown in practical modeling
		applications. Thus, estimating these quantities with $\hat{\sigma}^2$ and $s_n(\hat{\theta})$ respectively, a reasonable estimate for the variance of $n\log(\hat{\sigma}^2)$ can
		be found to be
		\begin{equation*}
			Var \left[ n\log(\hat{\sigma}^2) \right] \approx \frac{n}{\hat{\sigma}^4} s_n(\hat{\theta}) .
		\end{equation*}
		By the relation presented in (2.1), it is clear that this variance estimate is also suitable for $Var \left[ -2 \ell (\hat{\theta}  ) \right]$, and therefore
		likelihood-based information criteria as a whole that possess a constant penalty term. This estimator should converge to our previously derived value $2n$ in the case of a correctly
		specified model, as the sandwich estimator component will converge to the expected Fisher information used earlier, and the MLE $\hat{\sigma}^2$ should converge to the true
		parameter value $\sigma^2$. However, this sandwich estimator need not assume correct specification, and should be relatively robust to model misspecification.
		
		For the remainder of this work, this sandwich estimator will be referred to as $\widehat{Var}[GOF]$, such that
		\begin{equation*}
			\widehat{Var}[GOF] = \frac{n}{\hat{\sigma}^4} s_n(\hat{\theta}) .
		\end{equation*}

		We have established an asymptotic variance for the likelihood goodness-of-fit term in the case of a correctly specified normal linear regression model, and an
		estimator for this variance that does not assume the model is correctly specified. We will now synthesize these two developments to form a general goodness-of-fit procedure to
		test the hypothesis that a given normal linear regression model is correctly specified.

		Under the null hypothesis that a normal linear regression model is correctly specified, the estimator $\widehat{Var}[GOF]$ should be close to the theoretical value $2n$ for a
		sufficient sample size. We propose the use of the non-parametric bootstrap to obtain an empirical estimate for the sampling distribution of $\widehat{Var}[GOF]$. Once this empirical distribution
		has been obtained, the null hypothesis can be tested by observing whether a bootstrap interval for $Var[GOF]$ contains the theoretical value $2n$, as the approximation
		$Var[GOF] \approx 2n$ should hold for sufficient sample sizes under the null hypothesis. If a $100*(1-\alpha)$\% bootstrap confidence interval does not contain 
		$2n$, we reject the null hypothesis and conclude that the model is misspecified; however, if the interval does contain $2n$, we do not have sufficient evidence to reject
		the null hypothesis, and the proposed model does not demonstrate lack-of-fit.

		A full summary of the proposed procedure can be found below.
		\begin{algorithm*}[h]
			\caption*{$\bf{Algorithm}$ Bootstrap Goodness-of-Fit Test for a Normal Linear Regression Model}
			\begin{algorithmic}[1]
			  \Statex For test level $\alpha$, candidate normal linear model $M$, bootstrap iterations $B$, sample size $n$, and a null hypothesis that $M$
			  is adequately specified:
			  \State Resample, with replacement, outcomes with covariates to generate a bootstrap sample of size $n$.
			  \State Fit model $M$ to this bootstrap sample, and with this fitted model, calculate $\widehat{Var}[GOF]$
			  and record this statistic.
			  \State Repeat steps 1-2 $B$ times to generate an empirical bootstrap distribution for $\widehat{Var}[GOF]$.
			  \State Construct a $100*(1-\alpha)$\% bootstrap confidence interval for $Var[GOF]$.
			  \State If this interval does not contain $2n$, reject the null hypothesis at the $\alpha$ level. If it does contain
			  $2n$, the null hypothesis was not rejected and model $M$ does not appear to exhibit lack-of-fit. 
			\end{algorithmic}
		\end{algorithm*}

		This procedure can be used to assess the general hypothesis that a normal linear regression model is properly specified against the alternative that it displays lack-of-fit.
		Unlike other goodness-of-fit tests and procedures, this method does not test a specific assumption of linear regression such as normality or homoskedasticity, but rather
		all assumptions of normal linear regression. If any of them are violated, the property $Var[GOF] \approx 2n$ is not guaranteed to hold. This characteristic allows the test to detect
		many forms of misspecification from mean misspecification to distributional misspecification to heteroskedasticity.
		
		While the simulations in the following section will employ a percentile interval as the bootstrap confidence interval method of choice, one is not limited to using this method
		and may use any bootstrap interval they choose so long as it is theoretically justifiable. Additionally, the non-parametric bootstrap is employed in the algorithm detailed above to avoid further
		assumptions. Limited testing results suggest that the residual bootstrap may also work just as well in this procedure. However, pending further investigation, the non-parametric
		bootstrap is recommended at the current time. All simulations presented in this work will use the non-parametric bootstrap flavor of the procedure.


\section{Simulation Studies}

Four simulations will be employed to assess the efficacy of the bootstrap procedure developed in the previous section. In each simulation, an $n$ by 1 outcome vector $y = [y_1,...,y_n]'$ will
be generated for $i = 1,...,n$ according to
\begin{equation*}
	y_i = 2.0 + 2.0 x_{i1} + 2.0 x_{i2} + \epsilon_i , 
\end{equation*}
with the nature of the error term $\epsilon$ and the fitted candidate model varying between different simulations.

In the first simulation, each $\epsilon_i$ will be generated as $\epsilon_i \stackrel{iid}{\sim} N(0,4)$, and $x_{i1}$ and $x_{i2}$ are completely $iid$ covariates generated according to
a $Uniform(0,5)$ distribution. Note this formulation is one of a normal linear regression model with an intercept and an effect for each covariate present. With the generated data in hand, we will
proceed to fit a normal linear regression model that includes an intercept and effects for $x_{i1}$ and $x_{i2}$. This model is properly specified, and should not exhibit gross lack-of-fit.

In the second simulation scenario, each $\epsilon_i$ will be generated as $\epsilon_i \stackrel{iid}{\sim} N(0,4)$, and $x_{i1}$ and $x_{i2}$ are completely $iid$ covariates generated according to
a $Uniform(0,5)$ distribution. However, in this case the fitted model will be a normal linear regression model with an intercept and effect for only
$x_{i1}$. The covariate $x_{i2}$ will be omitted from the model, and could be viewed as a characteristic that affects the true generating process, but
was unobserved in data collection. This omission will induce mean misspecification into this fitted model as the mean structure will be underspecified.

In the third simulation scenario, each $\epsilon_i$ will be generated as $\epsilon_i \stackrel{iid}{\sim} N \left( 0,(2 + x_{i3})^2 \right)$, and $x_{i1}$, $x_{i2}$, and $x_{i3}$ are completely $iid$ covariates
generated according to a $Uniform(0,5)$ distribution. Note that heteroskedasticity is introduced into this model courtesy of $x_{i3}$ affecting the error variance. The fitted model for
this simulation will be a normal linear regression model that has the correct mean structure of an intercept with effects for $x_{i1}$ and $x_{i2}$, but will be fit assuming uniform variance.
This will result in unbiased estimates, and standard inference will be valid as well, as the heteroskedasticity present is related to a variable that isn't included in the model matrix; however,
these estimates are not guaranteed to be the most efficient.

In the final simulation scenario, each $\epsilon_i$ will be generated as $\epsilon_i \stackrel{iid}{\sim} N \left( 0,(2 + 0.5 x_{i2})^2 \right)$, and $x_{i1}$ and $x_{i2}$ are completely $iid$ covariates generated according to
a $Uniform(0,5)$ distribution. The fitted model for this simulation will be a normal linear regression model that has the correct mean structure of an intercept with effects for $x_{i1}$ and $x_{i2}$,
but will be fit assuming homoskedasticity. In this case the heteroskedasticity present is related to a covariate in the model matrix, and thus any normal linear model fit to the data will not be properly
specified if it assumes homoskedasticity, leading to potentially improper inference.

In each simulation, the bootstrap goodness-of-fit test, White test, and Breusch-Pagan test will each be performed at the $\alpha = .05$ level on the  fitted model. We expect the tests to roughly maintain their
Type~I error rates in the cases where the null hypothesis of each is true, and reveal the power of the test in the simulations where the null hypothesis is violated. After many simulation iterations, the
Type~I error rates or power of each test will be calculated based on the proportion of times each test rejected its null hypothesis. Each simulation will be performed for $n = 100, 500, 1000$ and $2500$, with $1000$
bootstrap iterations for all simulation. The simulation will be performed $1000$ times for each value of $n$.

Each simulation was performed in Julia, version 1.8.1 (\cite{Bezanson}). The code for the simulations can be found at \url{https://github.com/shkoeneman/GOF_manuscript/tree/main/scripts}. The results for the first simulation
can be found in the table below.

\begin{table}[H]
	\centering
	\small\addtolength{\tabcolsep}{-3pt}
	\setlength\extrarowheight{-3pt}
	\ttabbox[\FBwidth]
	{\caption{\label{tab:sim1_table}Simulation 1 Results - Type~I Error}}
	{
	\begin{tabular}{ c|c|c|c}
	$n$ & Bootstrap Test & White Test & Breusch-Pagan Test \\
	 \hline
	 100 & 0.094 & 0.071 & 0.059 \\
	 500 & 0.088 & 0.045 & 0.049 \\
	 1000 & 0.077 & 0.053 & 0.051 \\
	 2500 & 0.069 & 0.046 & 0.040 \\
	 \Xhline{3\arrayrulewidth}
	\end{tabular}
	}
\end{table}

The bootstrap test exhibits slight anti-conservative behavior across all values of $n$ with the empirically determined Type~I error rate exceeding the desired Type~I error rate
of $\alpha = 0.05$ in all scenarios. However, as $n$ increases, the empirical Type~I error rate appears to be decreasing towards the desired value. Additionally, the
behavior of the test is consistently anti-conservative for the scenarios shown. Thus, when using the procedure, one may keep in mind that the test may be more prone to
rejection than one might expect, particularly for small sample sizes before asymptotic properties of the bootstrap truly manifest. The White and Breusch-Pagan tests roughly maintain
the desired Type~I error rate across all values of $n$.

The results for the second simulation can be found in the table below.

\begin{table}[H]
	\centering
	\small\addtolength{\tabcolsep}{-3pt}
	\setlength\extrarowheight{-3pt}
	\ttabbox[\FBwidth]
	{\caption{\label{tab:sim1_table}Simulation 2 Results - Power/Type~I Error}}
	{
	\begin{tabular}{ c|c|c|c}
	$n$ & Bootstrap Test & White Test & Breusch-Pagan Test \\
	 \hline
	 100 & 0.549 & 0.050 & 0.043 \\
	 500 & 0.967 & 0.053 & 0.049 \\
	 1000 & 0.997 & 0.063 & 0.053 \\
	 2500 & 1.000 & 0.048 & 0.053 \\
	 \Xhline{3\arrayrulewidth}
	\end{tabular}
	}
\end{table}

The power of the bootstrap test in this scenario of mean misspecification is rather modest for low values of $n$ displayed in the table. However, power rises quickly to high levels
with the test appearing to possess near-perfect power as the value of $n$ approaches 2500. Thus, the bootstrap test appears to possess the ability to detect this mean
misspecification on account of an unobserved covariate, and the power to detect this misspecification rises as the sample size increases as one would expect.

In contrast, the White test and Breusch-Pagan test both seem to only maintain their Type~I error rates established in Simulation 3. This performance was to be expected
as the null hypotheses of these tests are not violated in this scenario. However, this simulation setting does demonstrate how the inability of a goodness-of-fit test to detect lack of
fit does not mean lack-of-fit is not present, as these two tests are limited in the scope of misspecification they are able to detect.

The results for the third simulation can be found in the table below.

\begin{table}[H]
	\centering
	\small\addtolength{\tabcolsep}{-3pt}
	\setlength\extrarowheight{-3pt}
	\ttabbox[\FBwidth]
	{\caption{\label{tab:sim1_table}Simulation 3 Results - Power/Type~I Error}}
	{
	\begin{tabular}{ c|c|c|c}
	$n$ & Bootstrap Test & White Test & Breusch-Pagan Test \\
	 \hline
	 100 & 0.108 & 0.057 & 0.057 \\
	 500 & 0.881 & 0.048 & 0.047 \\
	 1000 & 0.996 & 0.058 & 0.050 \\
	 2500 & 1.000 & 0.057 & 0.049 \\
	 \Xhline{3\arrayrulewidth}
	\end{tabular}
	}
\end{table}

The bootstrap test seems to only exhibit power in line with its empirical Type~I error rate displayed in Simulation 3 for $n = 100$, but the power of the test rises
rapidly, reaching near-perfect levels by the time $n = 1000$. In contrast, the White test and Breusch-Pagan test exhibit only power in line with their Type~I error
rates established in Simulation 3, and thus do not seem to be able to detect this form of heteroskedasticity, which is rather subtle and may only affect efficiency
of estimates as opposed to bias.

This limitation in detection is on account of how these tests are performed, as they rely on looking for heteroskedasticity that is induced by observed covariates.
When heteroskedasticity is induced by an unobserved covariate that is not used in the model, the White test and Breusch-Pagan test cannot ascertain that their
null hypotheses of homoskedasticity are violated. However, the bootstrap test is able to reject its more general null hypothesis.

The results for the fourth and final simulation can be found in the table below.

\begin{table}[H]
	\centering
	\small\addtolength{\tabcolsep}{-3pt}
	\setlength\extrarowheight{-3pt}
	\ttabbox[\FBwidth]
	{\caption{\label{tab:sim1_table}Simulation 4 Results - Power}}
	{
	\begin{tabular}{ c|c|c|c}
	$n$ & Bootstrap Test & White Test & Breusch-Pagan Test \\
	 \hline
	 100 & 0.040 & 0.462 & 0.674 \\
	 500 & 0.344 & 1.000 & 1.000 \\
	 1000 & 0.700 & 1.000 & 1.000 \\
	 2500 & 0.989 & 1.000 & 1.000 \\
	 \Xhline{3\arrayrulewidth}
	\end{tabular}
	}
\end{table}

In this scenario, the White test and Breusch-Pagan test both greatly outperform the bootstrap test in terms of power for lower values of $n$, with the bootstrap test
eventually matching the power of the other tests are $n = 2500$. This is likely due to the heteroskedasticity being induced by a covariate that is also in the 
proposed mean structure, the very case for which the White and Breusch-Pagan tests were designed. However, the bootstrap test is still able to detect this form of
misspecification in addition to other forms.

\section{Discussion and Conclusion}

In this work, we have developed a new bootstrap-based goodness-of-fit procedure to assess the goodness-of-fit of a fitted normal linear regression model. In order
to develop this test, we first derived an asymptotic variance for the goodness-of-fit term present in likelihood-based information criteria such as AIC under the assumption
that the fitted regression model is properly specified. The test functions by assessing whether the fitted model's estimated variance of this goodness-of-fit term conforms to the
theoretical value under proper specification. Simulation studies demonstrated the ability of this bootstrap test to detect a wide variety of violations of assumptions used in
normal linear regression. These violations include mean misspecification and violations to homoskedasticity, and this new procedure has the potential to serve as an omnibus
goodness-of-fit procedure for normal linear regression models.

This procedure has the potential to assist in model selection. For example, if one were to consider a normal linear regression framework to model a certain outcome, one may
choose to employ the bootstrap test on the largest possible model that one may consider, as if this model is misspecified, then all nested models will also be misspecified.
If the test does not reject its null hypothesis of proper specification, one could proceed with the normal linear regression framework and select a final model using
an information criteria such as AIC or BIC, or by employing another method such as cross validation.

This bootstrap procedure is not without its drawbacks. While the test is able to detect heteroskedasticity induced by unobserved covariates, this may be viewed as a detriment,
as this form of heteroskedasticity will not affect bias or invalidate classical likelihood inference, and will only affect efficiency of estimates. Thus, if the test rejects its
null hypothesis but one still would like  to move forward with a normal linear regression model, it is recommended to use robust estimates for the variance of estimated effects
to hedge against performing improper inference. Additionally, the test may be overpowered for real-life datasets with large sample sizes, resulting in rejection of the null hypothesis
when in fact a linear regression is a reasonable model. This is a feature common to many goodness-of-fit tests, and must be remembered when performing an analysis on observed data.

Potential future work involves finding similar procedures that can be used for other modeling frameworks, and finding further applications for the variance for
the goodness-of-fit term present in likelihood-based information criteria.

\section*{Appendix}

This Appendix will derive an exact variance for the goodness-of-fit term $-2 \ell (\hat{\theta} )$ in the case of a properly specified
linear model. We assume that this linear regression model has a full rank $n$ by $r$ design matrix $X$, outcome vector $y$, true parameters $\beta$ and $\sigma^2$,
and maximum likelihood estimates $\hat{\beta}$ and $\hat{\sigma}^2$. It can be shown that the estimate $\hat{\sigma}^2$ can be expressed as
\begin{equation*}
	\hat{\sigma}^2 =  \frac{1}{n} (y-X\hat{\beta})'(y-X\hat{\beta}) = \frac{1}{n} y'(I_n - X(X'X)^{-1}X')y ,
\end{equation*}
where $I_n$ is an $n$ by $n$ identity matrix. We let
\begin{equation*}
	A = \frac{1}{\sigma^2} (I_n - X(X'X)^{-1}X') 
\end{equation*}
such that we can construct the quadratic form
\begin{equation*}
	y'Ay = \frac{n \hat{\sigma}^2}{\sigma^2} .
\end{equation*}
Let $V = \sigma^2 I_n$ be the covariance matrix of the random multivariate normal outcome vector $y$, and note that $AV$ is idempotent. Let $\mu = X \beta$ be the
true mean of the outcome vector $y$. By the properties of quadratic forms of multivariate normal random variables, it can be seen that
\begin{equation*}
	y'Ay = \frac{n \hat{\sigma}^2}{\sigma^2} \sim \chi^2_{rank \left( A \right) } (\mu 'A\mu) ,
\end{equation*}
where $rank \left( A \right)$ is the degrees of freedom and $\mu 'A \mu$ is the non-centrality parameter of this chi-squared distribution.

Note that, as we assume $X$ is full rank, $rank(X) = r$. Furthermore, it can be seen that
\begin{equation*}
	\begin{split}
	tr\left[ X(X'X)^{-1}X' \right] & = tr \left[ X'X(X'X)^{-1} \right]  \\ 
	& = tr \left[ I_r \right] \\
	& = r .
	\end{split}
\end{equation*}
Thus, noting that $\sigma^2 A$ is idempotent, it can be seen that
\begin{equation*}
	\begin{split}
	tr \left[ \sigma^2 A \right] & = tr \left[ I_n - X(X'X)^{-1}X' \right]  \\ 
	& = n-r \\
	& = rank \left( \sigma^2 A \right) .
	\end{split}
\end{equation*}
Therefore, as $rank \left( \sigma^2 A \right) = rank \left( A \right)$ since $\sigma^2$ is a scalar constant, then $rank \left( A \right) = n-r$. This value will be the degrees of
freedom in the chi-squared distribution presented above.

Now, we observe the non-centrality parameter $\mu 'A \mu = (X \beta)'A(X \beta)$. Note that
\begin{equation*}
	\begin{split}
	A(X \beta) & = \frac{1}{\sigma^2} (I_n - X (X' X)^{-1} X')X\beta  \\ 
	& = \frac{1}{\sigma^2} (X \beta - X \beta) \\
	& = 0 .
	\end{split}
\end{equation*}
Thus, by extension $\mu 'A \mu = 0$, and this distribution is a central chi-squared distribution.

We have shown that
\begin{equation}
	\frac{n \hat{\sigma}^2}{\sigma^2} \sim \chi^2_{n-r} .
\end{equation}
Note that if we take the natural logarithm of the left-hand side and take its variance, we see that
\begin{equation}
	\begin{split}
		Var \left[ \log(\frac{n \hat{\sigma}^2}{\sigma^2}) \right] & = Var \left[ \log n + \log \hat{\sigma}^2 - \log \sigma^2 \right]  \\ 
		& = Var \left[ \log \hat{\sigma}^2 \right] \\
	\end{split}
\end{equation}
due to constant terms being irrelevant to the variance calculation. This enclosed form is very close to the $n \log \hat{\sigma}^2$ that is required
to determine the variance of the likelihood goodness of fit term $-2 \ell (\hat{\theta} )$ in the case of a normal linear regression model.
Thus, if we can calculate $Var \left[ \log \hat{\sigma}^2 \right]$, which is itself the variance
of the natural logarithm of a variable distributed as $\chi^2_{n-r}$, we can easily obtain $Var \left[ n \log \hat{\sigma}^2 \right]$.

Thus, we now derive the variance of the logarithm of a central chi-squared random variable. Let $W \sim \chi^2_{\nu}$, and let $Y = \log(W)$. Note that
the moment generating function of $Y$ can be expressed as
\begin{equation*}
	M_Y (t) = E \left[ e^{yt} \right] = E \left[ e^{t\log(w)} \right] = E \left[ e^{\log w^t} \right] = E \left[ w^t \right] .
\end{equation*}
Thus, the moment generating function of $Y$ evaluated at $t$ is the $t^{th}$ moment of $W$. The moments of $W$ will have the
closed form (Lancaster, 1969) of
\begin{equation*}
	E \left[ w^t \right] = 2^t \frac{\Gamma (t + \frac{\nu}{2})}{\Gamma (\frac{\nu}{2})} = E \left[ e^{yt} \right] = M_Y (t) .
\end{equation*}
We will use this relation to find the moments of $Y$, the logarithm of a central chi-squared random variable.

Taking the first derivative with respect to $t$, we see that
\begin{equation*}
	\frac{\partial M_Y(t)}{\partial t} = (2^t \log 2) \frac{\Gamma (t + \frac{\nu}{2})}{\Gamma (\frac{\nu}{2})} + 2^t \frac{\Gamma ' (t + \frac{\nu}{2})}{\Gamma (\frac{\nu}{2})} .
\end{equation*}
Note that in general $\frac{\Gamma ' (z)}{\Gamma (z)} = \psi^{(0)}(z)$, where $\psi^{(0)}(z)$ is the digamma function. With this relation in mind, we see that the first moment of
$Y$ can be calculated as
\begin{equation*}
	E \left[ Y^1 \right] =  \left. \frac{\partial M_Y(t)}{\partial t} \right|_{t=0} = \log 2 + \psi^{(0)} \left( \frac{\nu}{2} \right) .
\end{equation*}
Thus, we have a form for the first moment of the logarithm of a central chi-squared random variable.

Now moving on to calculating the second moment of $Y$, we take the next derivative of the moment generating function with respect to $t$ and see that
\begin{equation*}
	\begin{split}
		\frac{\partial^2 M_Y(t)}{\partial t^2} = (\log 2)^2 2^t  \frac{\Gamma (t + \frac{\nu}{2})}{\Gamma (\frac{\nu}{2})} + 2^{t+1} \log 2 \frac{\Gamma ' (t + \frac{\nu}{2})}{\Gamma (\frac{\nu}{2})} + \\ 
		2^t \frac{\Gamma '' (t + \frac{\nu}{2})}{\Gamma (\frac{\nu}{2})} .\\
	\end{split}
\end{equation*}
This above expression evaluated at $t = 0$ can be simplified by introducing the trigamma function $ \psi^{(1)}(z)$. Noting that the trigamma function is the derivative with respect
to $z$ of the digamma function $\psi^{(0)}(z) = \frac{\Gamma ' (z)}{\Gamma (z)}$, it can be seen that
\begin{equation*}
	\psi^{(1)}(z) = \frac{\partial \psi^{(0)}(z)}{\partial z} = \frac{\Gamma (z) \Gamma '' (z) - \Gamma '(z) \Gamma '(z)}{(\Gamma(z))^2} .
\end{equation*}
By re-arranging this relation, we see that
\begin{equation*}
	\Gamma ''(z) = \psi^{(1)}(z) \Gamma(z) + \frac{(\Gamma '(z))^2}{\Gamma(z)} .
\end{equation*}
Thus, moving to calculate the second moment of $Y$, we see that
\begin{equation*}
	E \left[ Y^2 \right] =  \left. \frac{\partial^2 M_Y(t)}{\partial t^2} \right|_{t=0} =
	(\log 2)^2 + \psi^{(0)} \left( \frac{\nu}{2} \right) 2 \log 2  + \frac{\Gamma '' (\frac{\nu}{2})}{\Gamma (\frac{\nu}{2})} .
\end{equation*}
Using the identity established above involving the trigamma function, we see that this moment can be expressed as
\begin{equation*}
	E \left[ Y^2 \right] =
	(\log 2)^2 + \psi^{(0)} \left( \frac{\nu}{2} \right) 2 \log 2  + \psi^{(1)} \left( \frac{\nu}{2} \right) + \left( \psi^{(0)} \left( \frac{\nu}{2} \right) \right)^2 .
\end{equation*}

Thus, with both the first and second moments of $Y$ in hand, we may calculate the variance. The variance of $Y$, the log of a central chi-squared random variable with
$\nu$ degrees of freedom, is found to be
\begin{equation*}
	Var \left[ Y \right ] = E \left[ Y^2 \right] - \left( E \left[ Y \right] \right)^2 = \psi^{(1)} \left( \frac{\nu}{2} \right) .
\end{equation*}
The variance will simply be the trigamma function evaluated at half of the degrees of freedom. This value can be calculated using software to approximate the trigamma function,
although it should be noted that this calculation may be unstable for certain values of the degrees of freedom.

Recalling the distributional result established in (A.1) and relation in (A.2), it is clear that
\begin{equation*}
	Var \left[ \log \hat{\sigma}^2 \right] = \psi^{(1)} \left( \frac{n-r}{2} \right)
\end{equation*}
based on the variance that has been derived above. Multiplying each side by $n^2$, we see that
\begin{equation*}
	Var \left[ n \log \hat{\sigma}^2 \right] = Var \left[ -2 \ell (\hat{\theta} ) \right] = n^2 \psi^{(1)} \left( \frac{n-r}{2} \right).
\end{equation*}
Thus, we have established the exact variance of the likelihood goodness-of-fit term in the case of a correctly specified linear model. This exact variance is in contrast to the
approximate variance found in the main body of this work.

However, while this variance is exact, it still involves an approximation as the trigamma function must be approximated. Additionally, it was found that
this variance provided no benefit when used in the bootstrap goodness-of-fit procedure developed in this work. Thus, this derivation is presented here only
for completeness and as a matter of theoretical interest.

\section*{Supplementary Materials}

There are no additional supplemental materials for this work.
\par
\section*{Acknowledgements}

We would like to acknowledge the University of Iowa for support during the development of this work.
\par


\bibhang=1.7pc
\bibsep=2pt
\fontsize{9}{14pt plus.8pt minus .6pt}\selectfont
\renewcommand\bibname{\large \bf References}
\expandafter\ifx\csname
natexlab\endcsname\relax\def\natexlab#1{#1}\fi
\expandafter\ifx\csname url\endcsname\relax
  \def\url#1{\texttt{#1}}\fi
\expandafter\ifx\csname urlprefix\endcsname\relax\def\urlprefix{URL}\fi


\begin{thebibliography}{}

\bibitem[Bezanson et al.(2017)]{Bezanson}
Bezanson, J., Edelman, A., Karpinski, S., and Shah, V.B. (2017).
 Julia: A Fresh Approach to Numerical Computing.
{\em SIAM Review}~{\bf 59}, 65--98.

\bibitem[Breusch and Pagan(1989)]{Breusch}
Breusch, T.S., and Pagan, A.R. (1989).
 A Simple Test for Heteroskedasticity and Random Coefficient Variation.
{\em Econometrica}~{\bf 47}, 1287--1294.

\bibitem[Cavanaugh and Neath(2019)]{Cavanaugh}
Cavanaugh, J.E., and Neath, A.A. (2019).
 The Akaike information criterion: Background, derivation, properties, application, interpretation, and refinements.
{\em WIREs Comput Stat}, 11:e1460.

\bibitem[Fisher(1922)]{Fisher}
Fisher, R.A. (1922).
 On the Mathematical Foundations of Theoretical Statistics.
{\em Phil. Trans. R. Soc. Lond. A}~{\bf 222}, 309--368.

\bibitem[Freedman(2006)]{Freedman}
Freedman, D.A. (2006).
 On The So-Called 'Huber Sandwich Estimator' and 'Robust Standard Errors'.
{\em The American Statistician}~{\bf 60}, 299--302.

\bibitem[Huber(1967)]{Huber}
Huber, P.J. (1967).
 The behavior of maximum likelihood estimates under nonstandard conditions.
{\em Proceedings of the Fifth Berkeley Symposium on Mathematical Statistics and Probability}~{\bf 5}, 221--233.

\bibitem[Kutner et al.(2005)]{Kutner}
Kutner, M.H., Nachtsheim, C.R., Neter, J., and Li, W. (2005).
 {\em Applied Linear Statistical Models}.
5th Edition. McGraw-Hill Irwin, New York.

\bibitem[Miles(2014)]{Miles}
Miles, J. (2014).
 Residual plot.
{\em Wiley StatsRef: Statistics Reference Online}.

\bibitem[Millar(2011)]{Millar}
Millar, R.B. (2011).
 {\em Maximum Likelihood Estimation and Inference: With Examples in R, SAS and ADMB.}
1st Edition. Wiley, Hoboken.

\bibitem[Rao(1965)]{Rao}
Rao, C.R. (1965).
 {\em Linear Statistical Inference and Its Applications.}
1st Edition. Wiley, Hoboken.

\bibitem[Schwarz(1978)]{Schwarz}
Schwarz, G.E. (1978).
 Estimating the dimension of a model.
{\em Annals of Statistics}~{\bf 22}, 461--464.

\bibitem[White(1980)]{White1980}
White, H. (1980).
 A Heteroskedasticity-Consistent Covariance Matrix Estimator and a Direct Test for Heteroskedasticity.
{\em Econometrica}~{\bf 48}, 817--838.

\bibitem[Waldman(1983)]{Waldman}
Waldman, D.M. (1983).
 A note on algebraic equivalence of White's test and a variation of the Godfrey/Breusch-Pagan test for heteroscedasticity.
{\em Economics Letters}~{\bf 13}, 197--200.



\end{thebibliography}


\vskip .65cm
\noindent
Thomas Jefferson University, Division of Biostatistics
\vskip 2pt
\noindent
E-mail: Scott.Koeneman@jefferson.edu
\vskip 2pt

\noindent
University of Iowa, Department of Biostatistics
\vskip 2pt
\noindent
E-mail: joe-cavanuagh@uiowa.edu

\end{document}